\documentstyle[aps,twocolumn,epsf,amssymb]{revtex}

\begin{document}

\draft

\title{Fluorescence from few electrons\thanks{Published in
\emph{Physical Review B}, \textbf{62} (2000) 13482--13489}}

\author{Jacques Arnaud}

\address{Mas Liron, F30440 Saint Martial, France}

\author{Laurent Chusseau}

\address{Centre d'\'Electronique et de Micro-opto\'electronique de
Montpellier \cite{UMR}, Universit\'e Montpellier II, F34095
Montpellier, France}

\author{Fabrice Philippe \cite{also}}

\address{D\'epartement de Math\'ematiques et Informatique
Appliqu\'ees, Universit\'e Paul Val\'ery, F34199 Montpellier, 
France}

\date{\today}

\maketitle


\begin{abstract}
Systems containing few Fermions (e.g., electrons) are of great current
interest.  Fluorescence occurs when electrons drop from one level to
another  without changing spin.  Only electron gases in a state
of equilibrium are considered.  When the system may exchange electrons
with a large reservoir, the electron-gas fluorescence is easily
obtained from the well-known Fermi-Dirac distribution.  But this is
not so when the number of electrons in the system is prevented from
varying, as is the case for isolated systems and for systems that are
in thermal contact with electrical insulators such as diamond.  Our
accurate expressions rest on the assumption that single-electron
energy levels are evenly spaced, and that energy coupling and spin
coupling between electrons are small.  These assumptions are shown to
be realistic for many systems.  Fluorescence from short, nearly
isolated, quantum wires is predicted to drop abruptly in the visible,
a result not predicted by the Fermi-Dirac distribution.  Our exact
formulas are based on restricted and unrestricted partitions of
integers.  The method is considerably simpler than the ones proposed
earlier, which are based on second quantization and contour
integration.
\end{abstract}

\pacs{05.30.Fk, 32.50.+d, 73.23.-b, 42.55.Ah}


\section{Introduction}
A number of remarkable experiments involving few electrons in
semiconductors and free-space, metal particles, and spin-1/2 atoms at
low temperatures, have been recently reported \cite{Zory},
\cite{Pell}.  Only electrons are considered below.  These collections
of electrons may be isolated or in thermal contact with the
environment, but, in any event, the number of particles is constant.
The Fermi-Dirac (FD) distribution holds when electrons may be freely
exchanged with a large reservoir (grand-canonical ensemble), but is
inaccurate for the systems considered.  The present paper provides
simple and accurate formulas for electron occupancy and fluorescence
for evenly-spaced single-electron energy levels.  Spontaneous emission
is supposed to be weak enough not to perturb importantly the system
state of equilibrium \cite{fai}.  Only vanishingly small Coulomb
interaction between electrons \cite{den} is considered.  The time
required for the system to reach equilibrium is not needed because
averagings are performed over unlimited time scales.  Quantum Optics
effects, such as resonance fluorescence or super-radiance, will not be
considered.

The assumption of evenly-spaced single-electron energy levels is not
as restrictive as one may think at first.  Consider indeed
one-dimensional devices such as the quantum wires employed in modern
laser diodes \cite{Zory}.  If the wire is uniform over its length and
the valence and conduction bands are parabolic in the energy-momentum
space, the energy spacing, $\epsilon$, between adjacent levels is not
a constant.  However, the variations of $\epsilon$ may be neglected
near the Fermi level as long as the temperature is not too high.  For
zero-band-gap semiconductors such as Pb$_{0.84}$Sn$_{0.16}$Te, the
energy-momentum relationship is linear rather than parabolic, and
$\epsilon$ is exactly a constant.  Level spacings in small irregular
metal particles (with a size on the order of $10~$nm) are nearly
uniform as a consequence of the mechanism of level ``repulsion''.  The
probability that adjacent levels be separated by $\epsilon$ is, for
the appropriate ensemble, of the form: $\epsilon ^{4} \exp(-\epsilon
^{2})$, a sharply peaked function of $\epsilon$ \cite{den}.  As is
well-known, the Landau levels that describe electron motion in uniform
magnetic fields \cite{Pell} are evenly spaced.  These levels are
highly degenerate, but the coupling between degenerate states
(expressing the drift of electrons through the magnetic field lines)
may be neglected over some period of time.  Likewise, two or
three-dimensional harmonic oscillators, modeling for example the
confinement of electrons in traps, exhibit degenerate evenly-spaced
levels.  Our approach may be generalized to degenerate levels.  It is
appropriate to mention also that the density of states of
(two-dimensional) quantum wells with parabolic bands is a constant
within a sub-band.  This implies that the energy-level spacing is
constant on the average, though not exactly.

The amount of light spontaneously emitted by electronic systems
depends on the optical-mode density of state, which is different for
free-space, low-dimensional structures, or photonic band gap
materials.  It is not the purpose of the present paper to discuss such
problems.  Because all comparisons are made at the same optical
wavelength, terms depending on wavelength only (essentially the
optical-mode density of state) may be dropped.  The quantity that we
calculate is the probability that the system exhibits an electron at
level $k$ and a hole (no electron) at level $k'$, with $k-k'=\hbar
\omega/\epsilon \equiv{d}$, where $\hbar$ denotes the Planck constant
divided by $2\pi$ and $\omega$ the angular optical frequency of
observation.  In pure semiconductors, electron-momentum conservation
entails that transitions may occur only symmetrically with respect to
the Fermi level, implying that $k = (1+d)/2$, $k' = (1-d)/2$, with odd
$d$, if $k=0$ labels the zero-temperature top electron.  We will
particularly consider the case where this condition holds.

Let us recall that, in Statistical Mechanics, isolated systems are
described by microcanonical ensembles, systems that may exchange
energy but not particles with a reservoir are described by canonical
ensembles, and systems that may exchange both energy and particles
with a reservoir are described by grand-canonical ensembles.  The
Fermi-Dirac (FD) distribution \cite{Kittel} is applicable to finite
systems only in the latter case.  That is, the FD distribution is
invalid for isolated systems and for systems that are in contact with
an electrical insulator such as diamond.  In the present context,
``finite'' (or ``small'', or ``mesoscopic'') means that $k_{B}T$,
where $T$ denotes the absolute electron gas temperature, is not
necessarily large compared with the adjacent-level energy spacings
$\epsilon$.  In grand canonical ensembles, the fluorescence is
proportional to the \emph{product} of the probabilities that the upper
level be occupied and that the lower level be empty.  In canonical
ensembles, it turns out that fluorescence is proportional to the
\emph{difference} between lower and upper levels occupancies.  This is
apparently a new result.

Exact formulas for level occupancy in finite single-spin systems with
evenly-spaced levels in contact with a heat bath have been reported
before \cite{den}, \cite{AJP}, \cite{Fab1}, \cite{Schon}, \cite{Fab2}.
Our method \cite{AJP} consists of first enumerating the microstates of
isolated systems.  Subsequent averaging provides expressions for the
canonical occupancies.  This
method is considerably simpler than the second-quantization methods
and integral-transformation formulas employed in \cite{den},
\cite{Schon}.  The present paper generalizes the results reported
in \cite{AJP} to account for the fact that some electrons may change
spin in the course of time.  Simple,
exact formulas for fluorescence are obtained for isolated systems and
canonical ensembles.  For arbitrary level energies and the
canonical ensemble, known recurrence formulas \cite{Borr}
are satisfied by our more
special, but explicit, forms.

The FD distribution is derived in Section 2 to set up the notation and
to recall why, in grand-canonical ensembles, spin is properly
accounted for by a two-fold degeneracy.  This method of accounting for
the electron spin, however, is invalid in microcanonical or canonical
ensembles because unpaired electrons may be in either one of two
distinguishable states.  Finite systems that exchange energy but not
particles with a reservoir are considered in Section 3 and 4.  In
Section 3, the electron spin is supposed to be strictly preserved and,
for simplicity, it is supposed that there are as many electrons with
spin up and with spin down.  Formulas valid for single-spin electrons
need only be multiplied by factors of 2 in that situation.  In Section
4, electrons are allowed to change spin in the course of time (but not
during a spontaneous emission event). For the sake of clarity, only
essential formulas relating to canonical ensembles are given in the
main text, detailed derivations and intermediate results being
relegated to Appendices A and B.

The purpose of Appendix\ \ref{sec.AppA} is to explain why the total
number $W$ of microstates of isolated systems is related to the
partitions of integers, for single-spin-state electrons.  It is shown
that the number $m(k)$ of microstates whose level $k$ is occupied is
simply related to $W$.  The number $m(k;k')$ of microstates whose
level $k$ is occupied and level $k'$ is empty is shown to be simply
related to $m(k)$, and thus to $W$.  Averaging, with the Boltzmann
factor as a weight, provides the corresponding formulas in the
canonical ensemble.  Appendix\ \ref{sec.AppB} explains how the
possibility that electrons may change spin in the course of time is
accounted for.

\section{The Fermi-Dirac Distribution}
\label{FD}

The probability $p$ that a system in thermal \emph{and} electrical
contact with a large medium contains $N^+$ electrons with spin up,
$N^-$ electrons with spin down, and energy $U$, is proportional to the
corresponding number of medium states (subscripts $m$) conveniently
written as $\exp \left( S_{m} ( N^+_{m}, N^-_{m}, U_{m} ) \right)$
with $S_{m}$ the system entropy.  If the system-medium contact is very
weak, energies as well as particle numbers add up.  A first-order
expansion of $S_{m}$ with respect to its arguments then gives
\begin{equation}
	p ( N^+ , N^- , U ) = C \exp(-\alpha N^+ -\alpha N^- -\beta U)
	\label{e1}
\end{equation}
where $C$ denotes a constant and
\begin{eqnarray}
	\beta & = & \frac{\partial S_{m}}{\partial U_{m}} \nonumber \\
	\alpha & \equiv & - \beta \mu = \frac{\partial S_{m}}{\partial
	N^+_{m}} = \frac{\partial S_{m}}{\partial N^-_{m}}\label{e2}
\end{eqnarray}
Here, $\beta\equiv1/k_{B}T$, where $T$ denotes the temperature, and
$\mu$ the Fermi level.  A single $\mu$ value occurs because the medium
behavior is the same for electrons of opposite spins.

It follows from Eq.\ (\ref{e1}) that the probabilities $p^{(0)}$ that
a nondegenerate state of energy $\epsilon_{k}$ is unoccupied,
$p^{(1)}$ that it is occupied by an electron of either spin, and
$p^{(2)}$ that it is occupied by two electrons, are respectively
\begin{eqnarray}
	p^{(0)} & = & C \nonumber \\
	p^{(1)} & = & C z \nonumber \\
	p^{(2)} & = & C z^2
	\label{e3}
\end{eqnarray}
with $z \equiv \exp ( - \alpha - \beta \epsilon_{k} )$.  Normalization
($p^{(0)} + 2 p^{(1)} + p^{(2)} = 1$) gives $C=1/(1+z)^2$.  The
occupancy (average number of electrons) $n_{FD} = 2 p^{(1)} + 2
p^{(2)}$ of level $k$ is therefore
\begin{eqnarray}
	n_{FD} ( q ; k ) & = & \frac{2}{z^{-1} + 1} \nonumber \\
	& = & \frac{2}{\exp ( \beta ( \epsilon_{k} - \mu ) ) + 1}
	\nonumber \\
	& = & \frac{2}{q^{\frac{1}{2} - k} + 1}
	\label{e4}
\end{eqnarray}
where $q \equiv \exp(-\beta)$ denotes the Boltzmann factor.  In the
last expression it is assumed that $\epsilon_{k} = k$, where $k$
denotes any relative integer.  The separation $\epsilon$ between
adjacent-level energies is taken as the energy unit, with a typical
value for $1~\mu$m--long quantum wires $\epsilon = 1\;$meV. In the
last expression in Eq.\ (\ref{e4}), $k=1$ labels the level just above
the top electron at $T = 0\;$K, and we have set $\mu = \frac{1}{2}$.
Note that for large $k$--values, $n_{FD} ( q ; k ) \approx 2
q^{k-1/2}$.

The average system energy added on top of the $T = 0\;$K energy is
obtained by summing up the occupancy over all levels, and subtracting
a similar sum for the $T = 0\;$K distribution.  The result is
\cite{AJP}
\begin{equation}
	r_{FD} (q) =\sum_{j=1,3\ldots}\frac {j}{q^{-j/2}+1}\label
	{rFD}
\end{equation}

In grand canonical ensembles, occupancies at different levels are
independent.  For a single spin state, this means that the probability
that level $k$ is occupied and level $k'$ is empty is the product of
level $k$ occupancy and [$1-$level $k'$ occupancy].  When the two
spin-states are considered, we obtain
\begin{equation}
	\label{e5}
	L_{FD} ( q ; k, k') = 2\frac{ n_{FD}( q, k)}{2} \left( 1 -
	\frac{n_{FD} ( q, k')}{2} \right)
\end{equation}
Fluorescence may indeed occur for 8 out of the 16 possibilities of
occupancy of levels $k$ and $k'$ (no electron, spin-up electron,
spin-down electron, or two electrons for each of the two levels).
Because occupancies are independent, the sum of the probabilities that
fluorescence events occur is indeed found to be given by Eq.\
(\ref{e5}).

If the electron-momentum conservation law is enforced, we have $k =
(1+d)/2$, $k' = (1-d)/2$, and the FD fluorescence in Eq.\ (\ref{e5})
reads after rearranging
\begin{equation}
	L_{FD} ( q ; d) = \frac{2}{{\left( q^{-d/2} + 1 \right)}^2}
	\label{e6}
\end{equation}
Thus, the fluorescence in the grand-canonical ensemble is given by a
simple function of temperature $T$ and angular optical frequency of
observation $\omega$.  We will see that the canonical ensemble
fluorescence is given by a simple series.

\section{Fluorescence without spin flip}

For single-spin electrons, the occupancy in isolated systems is given
by a simple formula reported in \cite{AJP}.  The proof, omitted in
\cite{AJP}, is given in Appendix\ \ref{sec.AppA} of the present paper
(see Eq.\ (\ref{eq:ni})).  If the energy added to the system is denoted
by $r$, the number $W(r)$ of configurations of the system is equal
here to the number $p(r)$ of \emph{partitions} of $r$.  Indeed,
microstates may be obtained by shifting electrons upward from their $T
= 0\;$K locations by non-increasing steps that sum up to $r$.  Let us
recall that a partition of $r$ is a set of positive integers summing
up to $r$.  For example $(2,1,1)$ is a partition of $4$.  The number
$p(4)$ of partitions of $4$ equals $5$.  By convention, $p(0)=1$ and
$p(r)=0$ if $r<0$.

To illustrate the difference existing between the exact result and the
Fermi-Dirac distribution, let us note that, for any microstate, the
energy separation between the top electron and the lowest hole cannot
exceed $r\epsilon $, where $\epsilon$ denotes as before the
adjacent-level energy spacing.  Accordingly, the fluorescence drops
abruptly at an angular optical frequency $\omega = r\epsilon /\hbar$.
When $r\gg 1$, a system temperature $T$ may be defined \cite{AJP}:
$k_{B}T\approx \epsilon \sqrt{6r}/\pi$.  As an example,
room-temperature isolated systems with $\epsilon=1$ meV should not
emit visible light according to the exact formula, while some faint
light is expected according to the FD distribution.

In the present section, it is supposed that the electron spins are
preserved in the course of time, and that there are as many electrons
with spin up and spin down, for simplicity.  It then suffices to
multiply the expressions for single-spin electron occupancy, average
energy, and fluorescence, given in Appendix\ \ref{sec.AppA}, by
factors of 2.

The occupancy reads, according to Eq.\ (\ref{eq:A12})
\begin{equation}
	\label{e8}
	n_{u}(q;k) = - 2\sum_{j=1,2\ldots}(-1)^{j} q^{j k + j (j -
	1)/2}
\end{equation}
Note that, for large $k$--values, $n_{u}(q;k) \approx 2 q^k$ so that
$n_{u}(q;k)/n_{FD}(q;k)=\sqrt{q}$, if the expression for $n_{FD}(q;k)$
in Eq.\ (\ref{e4}) is used.

The average added energy reads \cite{AJP}
\begin{equation}
	\label{ru}
	r_{u} (q)= 2 \sum_{j=1,2\ldots}\frac{j}{q^{-j}-1}
\end{equation}

We first compare in Fig.\ \ref{fig:1}a the occupancy in isolated
systems with the FD occupancy.  The former is obtained by multiplying
$n_{i}(r;k)$ given in Eq.\ (\ref{eq:ni}) by a factor of 2 to account
for the two spin states.  The FD occupancy is given in Eq.\
(\ref{e4}), with $q$ expressed in terms of the average energy $r$ with
the help of Eq.\ (\ref{rFD}).  Note that, below 0.1, isolated-system
occupancies are \emph{smaller} than FD occupancies.

Consider next the case where the system is in contact with a heat
bath.  The ratio of the exact occupancy in Eq.\ (\ref{e8}) (where $q$
is expressed in terms of the average energy $r$ with the help of Eq.\
(\ref{ru})) and the FD occupancy (Eqs.\ (\ref{e4}), (\ref{rFD})), is
represented in Fig.\ \ref{fig:1}b as a function of the FD occupancy
for various values of the average energy $r$ (namely, $r=6$, 60, and
600).  Below 0.1, the exact occupancy \emph{exceeds} the FD occupancy.
Figure \ref {fig:2}a shows that, when the comparison is made at equal
temperatures $T$ (instead of equal average energies), the opposite
occurs.  In Fig.\ \ref{fig:2}, we have chosen to represent the
occupancy ratios as functions of the level number $k$ (referred to the
Fermi level) instead of the FD occupancy.

To evaluate fluorescence, we need to know the number $m(r;k;k')$ of
microstates of added energy $r$ having an electron at level $k$ and
none at level $k'=k-d$, where $d \equiv \hbar \omega/\epsilon $.
Appendix\ \ref{sec.AppA} shows that this quantity is easily
expressible in terms of the numbers $m(r;k)$ of microstates having an
electron at level $k$ (irrespectively of other state occupancies).
Averaging the result over $r$ with $q^r$ as a weight to account for
energy fluctuations in the presence of a heat bath, the fluorescence
is found to be (see Eq.\ (\ref{L}))
\begin{equation}
	\label{e9}
	L_{u}(q;k,k') = \frac{n_{u}(q;k') - n_{u}(q;k)}{q^{-d}-1}
\end{equation}
where $n_{u}(q;k)$ is given in Eq.\ (\ref {e8}).

If the law of electron-momentum conservation is enforced, the
fluorescence reads
\begin{equation}
	\label{Lu}
	L_{u}(q;d)= L_{u}(q;\frac{1+d}{2},\frac{1-d}{2})
\end{equation}

\begin{figure}
     \centering
     \epsfbox{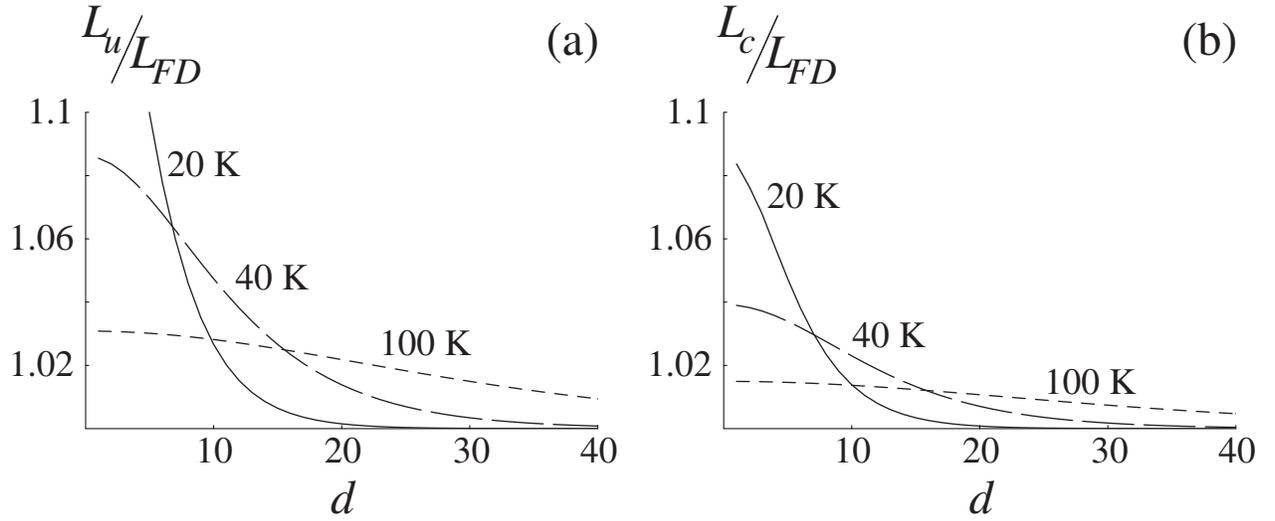}
     \caption{Ratio of spontaneously-emitted light power
     (fluorescence) from 1~$\mu$m-long quantum wires in contact with
     diamond (canonical ensemble) and copper (grand-canonical
     ensemble), respectively, as a function of $d \equiv \hbar \omega
     / \epsilon$.  The parameter is the electron gas absolute
     temperature $T$.  a) Spin flip is not allowed.  b) Spin flip is
     allowed.  }
     \label{fig:3}
\end{figure}

Recall that $q \equiv \exp(- \epsilon/ k_{B} T)$ (where $\epsilon$ is
typically $1\;$meV and room temperature corresponds to $k_{B}T =
26\;$meV), and $d \equiv \hbar \omega/\epsilon$.  The fluorescence
ratio: $L_{u}(q;d)/L_{FD}(q;d)$ according to Eq.\ (\ref{e6}) and Eq.\
(\ref{Lu}), is represented in Fig.\ \ref{fig:3}a as a function of $d$
for different temperatures.  It is interesting that canonical and
grand-canonical fluorescences almost coincide at small wavelengths
even though the occupancies are quite different in that limit.

\section{Fluorescence with spin flip}

Electron spins are now allowed to vary in the course of time (but not
during a fluorescence event).  The numbers $N^+$ and $N^-$ of
electrons with spin up and spin down, respectively, may fluctuate, but
their sum $N^+ + N^- \equiv N$ remains constant if the system is
isolated or in contact with an electrically-insulating heat sink such
as diamond.  The occupancy and fluorescence for coupled-spin electrons
is derived from previous expressions through a succession of
partitionings and averagings.  Because the details are lengthy, they
are relegated to Appendix\ \ref{sec.AppB}. Remarkably, many summations can be
performed in closed form so that the final result is simple.

The occupancy reads
\begin{equation}
	n_{c}(q;k) = \frac{\sum_{j= - \infty}^{+ \infty}q^{j^2}
	n_{u}(q;k+j)}{\sum_{j=- \infty}^{+ \infty}q^{j^2}}
	\label{e10}
\end{equation}
where $n_{u}(q;k)$ is given in Eq.\ (\ref{e8}).  Comparisons with the
FD distributions are exemplified in Figs.\ \ref{fig:1}d and
\ref{fig:2}b.  With the help of theta functions \cite{Berndt} the
average added energy may be written as a simple sum (see Eq.\
(\ref{eq:B5}))
\begin{equation}
	\label{e11}
	r_{c} (q) =2 \sum_{j=1,2\ldots}\left(
	\frac{j}{q^{-j} -1} + \frac{(-1)^j j}{q^j - q^{-j}} \right)
\end{equation}

The fluorescence reads (see Eqs.\ (\ref{A13}) and (\ref{eq:B6}))
\begin{equation}
	\label{e12}
	L_{c}(q;d) = \frac{\sum_{j = -\infty}^{+\infty}q^{j^2}
	L_{u}(q;j+\frac{1+d}{2},j+\frac{1-d}{2})} {\sum_{j=
	-\infty}^{+\infty}q^{j^2}}
\end{equation}
where $L_{u}(q;k,k')$ is given in Eq.\ (\ref{e9}).  Fluorescence is
illustrated in Fig.\ \ref{fig:3}b.  Note that the exact result is
closer to the FD result when electrons are allowed to change spin in
the course of time.

\section{Conclusion}
When a (possibly small) system is in thermal and electrical contact
with a large medium such as a piece of copper, the average number of
electrons occupying some energy level (occupancy) is twice the value
given by Fermi-Dirac (FD) formula.  The fluorescence (light
spontaneously emitted without electron-spin flip), defined in terms of
the probability that an electron at level $k$ may drop to level
$k-\hbar\omega/\epsilon $ ($\omega$ denotes the angular optical
frequency of observation), is also easily obtained.  But when the
system is isolated, or in thermal contact with an electrical
insulator, electron occupancies are given by different expressions.
Because modern electronics often employ short quantum wires supported
by diamond heat sinks, it is important to have at our disposal precise
expressions for occupancy and fluorescence in such situations.  The
expressions obtained in this paper were illustrated by comparison with
the FD results.  We considered the case where the electron spins are
strictly maintained in the course of time (Section 3), and the case
where spin-flip is allowed (Section 4).  We found, for example, that
small FD occupancies should be multiplied by approximately $\exp ( -
\epsilon / 2 k_{B} T ) \approx 0.22$ if $\epsilon = 1\;$meV and
$T=4\;$K, a factor that differs significantly from unity.  But,
unexpectedly, the fluorescence turns out to be given rather accurately
by the FD distribution.

Our mathematical approach is based on a direct enumeration of the
microstates, and the results are expressed in terms of the number of
partitions of integers.  This method is considerably simpler than
those previously reported for similar models, both conceptually and
algebraically.  A computer simulation has given results that are in
very good agreement with the analytical formulas reported in this
paper.

It is our intention to report in the future analytical and numerical
results relating to mesoscopic laser-diode light fluctuations.  A
preliminary step consists of considering single-mode cavities
incorporating the electron gas at thermal equilibrium, with one or two
bands of states (for a single band, see Appendix\ \ref{sec.AppA} of
the present paper).  The intraband Auger effect and the stimulated
transitions may be introduced at that stage.  Next, the probability
that low-lying electrons be promoted to high-lying levels by the
action of a (quiet or fluctuating) pump, and the probability that
light quanta be absorbed, are introduced.  At low power, our
simulation gives output light fluctuations that agree very well with
elementary laser-noise theory predictions \cite{chuss}.  At high
power, new effects (temperature fluctuations, spectral-hole burning,
statistical fluctuations of the optical gain $\ldots$) occur that are
difficult to handle analytically \cite {Arn}.  The analytical formulas
reported in this paper are helpfull to assess the accuracy of the
simulation in special situations.

\appendix

\section{Occupancy and fluorescence for single-spin-state electrons.
Arbitrary energy levels}
\label{sec.AppA}

We are only concerned in the main text with one-electron energy levels
$\{ \epsilon_{k} \} = \mathbb{Z}$.  Rigorous occupancy
formulas are obtained in the present Appendix by considering first
arbitrary $\{ \epsilon_{k} \}$.  Eventually the number of levels is
allowed to go to infinity.

Consider an isolated system whose nondegenerate one-electron level
energies are, in increasing order, $\epsilon_{1}$, $\epsilon_{2}$,
\ldots $\epsilon_{k}$, \ldots $\epsilon_{B}$, with $N \le B$
single-spin electrons.  According to the Pauli principle each level
may be occupied by only $0$ or $1$ electron.  The system energy $U$ is
therefore the sum of $N$ of the $\epsilon_{k}$.  The purpose of this
Appendix is to evaluate:

\begin{enumerate}

	\item The number $W(N,U)$ of possible ways of obtaining some
	given $U$ (number of microstates, or ``statistical weight'').

	\item The number $m(N,U;k)$ of microstates whose level $k$ is
	occupied.  The occupancy $n(N,U;k)$ of level $k$ is defined as
	$m(N,U;k)/W(N,U)$.

	\item The number $m(N,U;k;k')$ of microstates whose level $k$
	is occupied and level $k'$ is empty.  The fluorescence
	$L(N,U;k;k')$ emitted by electrons dropping from level $k$ to
	level $k'$ is defined as $m(N,U;k;k')/W(N,U)$.

\end{enumerate}

These evaluations will be presented in reversed order.  Let us first
relate the number $m(N,U;k;k')$ of microstates whose level $k$ is
occupied and level $k'$ is empty, to the numbers $m(N,U;k)$ defined
above.  For each microstate, let the electron at level $k$ be
transferred to the lower empty level $k'$.  The number $N$ of
electrons is unaffected but the total energy gets reduced from $U$ to
$U-d$ where $d=\epsilon_{k}-\epsilon_{k'}>0$, and the roles of $k$ and
$k'$ are reversed.  The following equality
\begin{equation}
	m(N,U;k;k') = m(N,U-d;k';k)
	\label{m3}
\end{equation}
therefore holds.  Now notice that,
\begin{eqnarray}
	\lefteqn{m(N,U;k) - m(N,U;k;k') =} \nonumber \\
	& & m(N,U;k') - m(N,U;k';k)
	\label{m4}
\end{eqnarray}
because the two sides of the above equation count microstates whose
levels $k$ and $k'$ are \emph{both} occupied.  When the expression in
Eq.\ (\ref{m4}) is introduced in Eq.\ (\ref {m3}) iteration gives a recurrence
relation for $m(N,U;k;k')$ that reads
\begin{eqnarray}
	\lefteqn{m(N,U;k;k') =} \nonumber \\
	& & \sum_{j=1,2\ldots } \left( m(N,U-jd;k') - m(N,U-jd;k)
	\right)
	\label{m6}
\end{eqnarray}
The above series terminates when the total energy vanishes, i.e., when
$jd$ exceeds $U$.

Consider next the $m(N,U;k)$ microstates whose levels $k$ are
occupied, and remove these electrons.  The same number of microstates
is obtained, with $N-1$ electrons, total energy $U-\epsilon_{k}$, and
no electron at level $k$.  The number of these new microstates may be
written as the difference between the \emph{total} number of
microstates and the number of microstates whose level $k$ is
\emph{occupied}.  We have therefore the identity
\begin{equation}\label{m}
	m(N,U;k) = W(N-1,U-\epsilon_{k}) - m(N-1,U-\epsilon_{k};k)
\end{equation}
After a sufficient number of iterations, either the energy or the
number of electrons becomes negative and the last term vanishes.  The
quantity $m(N,U;k)$ may therefore be written as a finite sum
\begin{equation}\label{m1}
	m(N,U;k) = - \sum_{j=1,2\ldots }(-1)^{j} W(N-j,U-j\epsilon _{k})
\end{equation}
The series terminates when $j$ exceeds either $N$ or $U /
\epsilon_{k}$.

We have the obvious identity (the number of occupied states for the
whole set of microstates being written in two different manners)
\begin{equation}
	N W(N,U)=\sum_{k \geq 1}m(N,U;k)
	\label{A6}
\end{equation}

A system in contact with a heat bath at temperature $T$ is described
by the canonical ensemble.  Let us define as in the main text $q
\equiv \exp ( - \beta )$, where $\beta \equiv 1 / k_{B} T$.  The
so-called partition function $Z(N,q)$ is the sum over $U$ of $q^{U}
W(N,U)$, and the average energy is $( q / Z ) \partial Z(N,q) /
\partial q$.  When both sides of Eq.\ (\ref{A6}) are multiplied by
$q^U$ and summed over $U$ we obtain, using Eq.\ (\ref{m1})
\begin{eqnarray}
	Z(N,q) & = & \frac{1}{N} \sum_{k=1,2\ldots} \sum_{U}
	q^{U} m(N,U;k) \nonumber \\
	& = & - \frac{1}{N} \sum_{j=1,2\ldots} (-1)^j
	\sum_{k=1,2\ldots} q^{j \epsilon_{k}} \nonumber \\
	& & \times \sum_{U} q^{U - j \epsilon_{k}} W(N-j,U-j
	\epsilon_{k}) \nonumber \\
	& = & - \frac{1}{N} \sum_{j=1,2\ldots N} (-1)^j Z(1,q^j)
	Z(N-j,q)
     \label{eq:A7}
\end{eqnarray}

Indeed, for a single electron ($N = 1$), $U$ may only take one of the
$\epsilon_{k}$ values and the statistical weight $W$ is unity.  It
follows that $Z(1,q)$ is the sum over $k$ from 1 to $B$ of
$q^{\epsilon_{k}}$.  Note that $Z(0,q)=1$.  Equation (\ref{eq:A7}) was
obtained earlier \cite{Borr2} from a less direct proof.

The occupancy (average number of electrons) $n(q;k)$ of level $k$, is
equal to the sum over $U$ of $q^{U} m(N,U;k)$, divided by $Z(N,q)$,
where $m(N,U;k)$ is given in Eq.\ (\ref{m1}) and $Z(N,q)$ is defined
in Eq.\ (\ref{eq:A7}) from a recurrence relation.  Thus
\begin{equation}\label{nu1}
	n(q;k) = \frac{- \sum_{j=1,2\ldots}(-1)^{j}
	q^{j\epsilon_{k}}Z(N-j,q)}{Z(N,q)}
\end{equation}
Expression (\ref{nu1}) was reported before \cite{Borr}.

The probability that level $k$ be occupied and that level $k'$ be
empty at temperature $T$ is similarly obtained by summing $q^{U}
m(N,U;k;k')$ over $U$, and dividing the result by $Z(N,q)$, where
$m(N,U;k;k')$ is given in Eq.\ (\ref{m6}).  The result of the
summation may be expressed as the \emph{difference} of the lower and
upper occupancies, according to
\begin{equation}\label{L}
	L(q;k,k') = \frac{n(q;k') -
	n(q;k)}{q^{\epsilon_{k'}-\epsilon_{k}}-1}
\end{equation}
where the occupancy $n(q;k)$ is given in Eq.\ (\ref{nu1}).  Thus the
fluorescence is equal to the difference between the occupancies at
$k'$ and $k$ multiplied by the Bose function.

Let us now specialize the above formulas for the case where
$\epsilon_{k}=k$, $k=1\dots B$.  Considering the displacement of the
$N$ electrons from their least-energy locations ($k=1$ to $N$)
beginning to the one on top, we observe that $W(N,U)$ is the number
$p(P;N,r)$ of partitions of the \emph{added energy} $r \equiv U - N (
N+1 ) / 2$ into at most $N$ parts, none of them exceeding $P\equiv
B-N$.  Note that the numbers $p(P;N,r)$ may be obtained from a
recurrence relation \cite{Andrew}, $p(P;N,r)-p(P;N-1,r)=p(P-1;N,r-N)$.
Let us now change slightly our notation, letting $k=0$ denote the top
electron in the least energy configuration, and let us employ the
added energy $r$ instead of the total energy $U$ as an argument.
Equation (\ref{m1}) reads
\begin{eqnarray}
	\lefteqn{m(N,r;k) = - \sum_{j=1,2\ldots }(-1)^{j}} \nonumber \\
	&& \times p ( P+j ; N-j , r-jk- j(j-1)/2 )
	\label{A10}
\end{eqnarray}

If $r$ does not exceed $N$ and $P$, it is intuitive that $p(P;N,r) =
p(r)$, where $p(r)$ denotes the number of unrestricted partitions of
$r$.  Equation (\ref{A10}) then simplifies to
\begin{equation}\label{A11}
	m(r;k) = - \sum_{j=1,2\ldots }(-1)^{j} p( r - j k -
	j(j-1)/2 )
\end{equation}
This expression was reported (for the first time to our knowledge), in
\cite{AJP}.  If $N$ and $P$ are infinite ($\{ \epsilon_{k} \} =
\mathbb{Z}$), Eq.\ (\ref{A11}) holds for any finite value of $r$ and
the corresponding single-spin-state occupancy of an isolated system is
\begin{equation}
     n_{i}(r;k) = m(r;k) / p(r)
     \label{eq:ni}
\end{equation}
Averaging the numerator and denominator of above expression with $q^r$
as a weight with $r$ from 0 to $\infty$, gives the canonical occupancy
\cite{den}
\begin{equation}
     n(q;k) = - \sum_{j=1,2\ldots}(-1)^{j} q^{jk+j(j-1)/2}
     \label{eq:A12}
\end{equation}

Finally the expression in Eq.\ (\ref{L}) simplifies in the present
situation to
\begin{equation}\label{A13}
	L(q;k,k') = \frac{n(q;k') - n(q;k)}{q^{k'-k}-1}
\end{equation}

We have set up a Monte Carlo simulation program that enables us to
recover previous analytical expressions.  For the case of isolated
systems, a constant probability per unit time is ascribed to
level-changing events that preserve energy.  The system eventually
reaches a state of equilibrium with an electron distribution very
close to the one derived from previous recurrence formulas.  The
Fermi-Dirac distribution is obtained in the limit of large $B$-values,
with temperatures and Fermi levels that depend on the energy given
initially to the system.  Our computer program may handle
single-electron level distributions that could be difficult to analyze
theoretically (for example, two bands of states).

When the system is in thermal contact with a heat bath, electrons at
level $k$ are ascribed a probability per unit time, $p$, of being
promoted to level $k+1$ and a probability, $qp$, of being demoted to
level $k-1$, provided these levels are empty.  Strictly speaking,
these prescriptions rest on an Einstein-type model of solids that
supposes that the atoms are vibrating at frequency $\omega_{phonon} =
\epsilon /\hbar$, where $\epsilon$ denotes as before the electronic
level spacing.  But the details of the thermalization model turn out
to be rather unimportant.  The computer program enabled us to
reproduce the theoretical results with great accuracy.  For example,
when $B=100$, $\epsilon=1$meV, and $T=$100K, the numerical
distribution fits the Fermi-Dirac distribution with a discrepancy not
exceeding 0.2\%.

When the electron gas is enclosed in a single-mode cavity, the
probability that the cavity contains $m$ light quanta is proportional
to $W(r-md)$, where $W(r)$ denotes the statistical weight of the
electron gas for an added energy $r$, see Appendix\ \ref{sec.AppB} of
\cite{AJP}.  If, initially, only the highest levels are occupied, we
obtain exactly, from the recurrence relation satisfied by $p(P;N,r)$,
${\mathrm{variance}}(m)/{\mathrm{average}}(m)=(B+1)/6$.

\section{Occupancy and fluorescence for two-spin-states electrons}
\label{sec.AppB}

\begin{figure}
     \centering
     \epsfbox{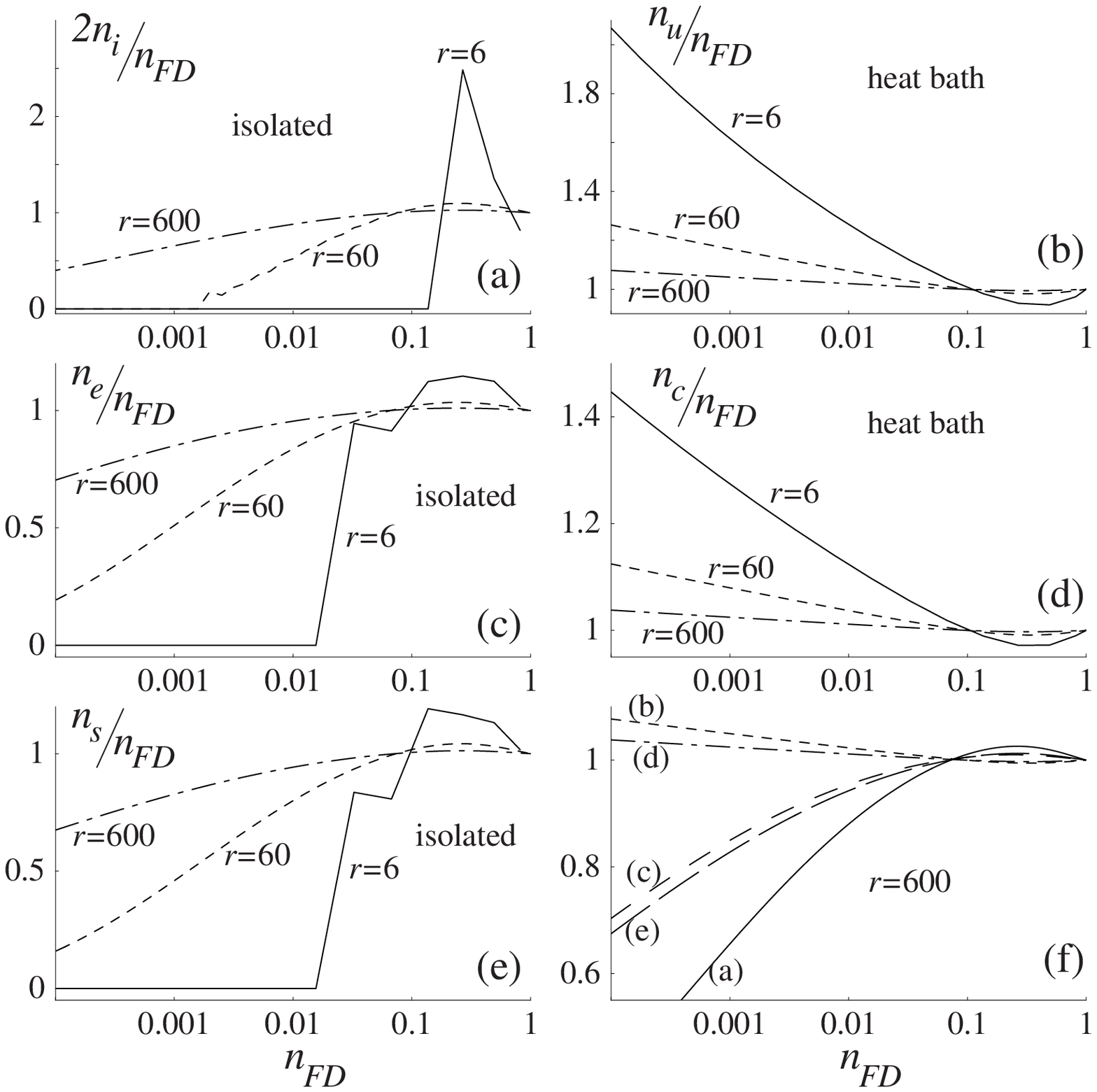}
     \caption{Ratio of exact and Fermi-Dirac ($n_{FD}$) occupancies
     plotted as functions of $n_{FD}$ at equal values of the average
     added energy $r$.  The isolated single-spin-state electron
     occupancy in a) was multiplied by 2 to account for the two spin
     states.  In a) and b) spin flip is not allowed.  In c) electrons
     of different spins may exchange energy but spin flip is not
     allowed.  Cases d) and e) are the same as cases b) and c)
     respectively, except that spin flip is allowed.  f) collects
     previous results for an average added energy $r=600$.}
     \label{fig:1}
\end{figure}

\begin{figure}
     \centering
     \epsfbox{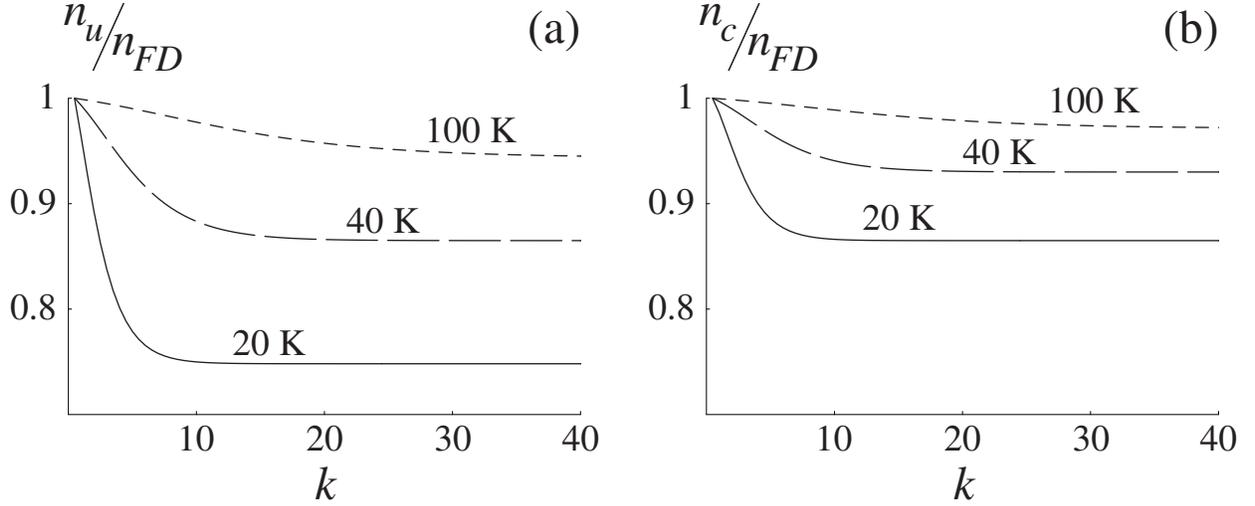}
     \caption{This figure is similar to Fig.\ \ref{fig:1}, but the
     comparison between canonical and grand-canonical ensembles is made
     at equal temperatures rather than at equal average energies.  The
     energy level spacing is supposed to be 1~meV, typical of 1~$\mu$m
     long quantum wires.  The parameter is the electron gas absolute
     temperature $T$.  a) Spin flip is not allowed.  b) Spin flip is
     allowed.  }
     \label{fig:2}
\end{figure}

In the present Appendix we restrict ourselves to energy levels
$\epsilon_{k} = k$, with $k=1,2\ldots$, the origin of the energy being
set at $k=0$.  Electrons are allowed to change spin in the course of
time.  We first consider an isolated system with constant numbers of
spin-up and spin-down electrons ($N^{+}$ and $N^{-}$, respectively),
electrons of different spins being allowed to exchange energy.  Next,
spin-flip is allowed.  Averaging, with the Boltzmann factor as a
weight, provides occupancies for the case where the systems are in
contact with a heat bath.  Occupancies in these various situations are
illustrated in Fig.\ \ref{fig:1}, again by comparison with the FD
distribution.

Consider first an isolated system with $N^{+}$ spin-up electrons and
$N^{-}$ spin-down electrons, and suppose that the two sub-systems may
exchange energy but that spin flip is not allowed.  Setting
$N^{+}+N^{-} \equiv 2 N$ and $N^{+}-N^{-} \equiv 2 n$, the system
least energy is
\begin{eqnarray}
     U_{0} & = & \frac{N^{+} ( N^{+}+1 )}{2} + \frac{N^{-} ( N^{-}+1
     )}{2} \nonumber \\
     & = & N(N+1)+n^2
     \label{eq:B1}
\end{eqnarray}

If $r \ge n^2$ denotes the energy added to the system on top of
$N(N+1)$, the remaining energy $s=r-n^2$ splits into $r_{1}$ in
sub-system 1 and $r_{2}=s-r_{1}$ in sub-system 2.

We have shown in Appendix\ \ref{sec.AppA} that, for the case presently
considered and in the limit $N \to \infty$, the number of microstates
for single-spin electrons is the number of partitions $p(r)$ of the
excess energy $r$.  The number of microstates relating to one
particular splitting of $s$ is therefore $p(r_{1}) p(r_{2})$.
Accordingly, occupancies are obtained by averaging single-spin state
occupancies shifted by $\pm n$, with a probability law proportional to
$p(r_{1}) p(s-r_{1})$ with $r_{1}$ running from 0 to $s$.
\begin{eqnarray}
     \lefteqn{n_{e}(r;k) =} \nonumber \\
     && \frac{\sum_{r_{1}} \left( m(r_{1};k-n) p(s-r_{1}) + m(s-r_{1};k+n)
     p(r_{1}) \right)}{\sum_{r_{1}}p(s-r_{1}) p(r_{1})} \nonumber \\
     &&
     \label{eq:B2}
\end{eqnarray}
where $m(r;k)$ is given by Eq.\ (\ref{A11}).  Figure\ \ref{fig:1}c
compares $n_{e}(r;k)$ in Eq.\ (\ref{eq:B2}) to $n_{FD}$ in Eqs.\
(\ref{e4}), (\ref{rFD}) for the case where $n=0$ ($N^{+}=N^{-}$ or
$r=s$) and various values of the added energy.

When spin flip is allowed, $N^{+}+N^{-}=2 N$ remains fixed, but $n
\equiv (N^{+}-N^{-}) / 2$ may take any value that does not exceed
$\sqrt{r}$, where $r$ denotes as before the energy added on top of
$N(N+1)$.  If $r=6$ for example, five values of $n$ are permitted,
namely $n = 0$, $n = \pm 1$ and $n = \pm 2$.  It thus suffices to sum
the numerator and denominator of Eq.\ (\ref{eq:B2}) over permissible
values of $n$.  The occupancy reads
\begin{eqnarray}
     \lefteqn{n_{s}(r;k) =} \nonumber \\
     && \frac{\sum_{n}\sum_{r_{1}}\left( m(r_{1};k-n) p(s-r_{1}) +
     m(s-r_{1};k+n) p(r_{1}) \right)}{\sum_{n}\sum_{r_{1}}p(s-r_{1})
     p(r_{1})} \nonumber \\
     &&
     \label{eq:B3}
\end{eqnarray}
where the sum over $r_{1}$ is from 0 to $s=r-n^2$, and $m(r;k)$ is
given Eq.\ (\ref{A11}).  The two terms in the numerator give equal
contributions.  Figure\ \ref{fig:1}e compares $n_{s}(r;k)$ as given in
Eq.\ (\ref{eq:B3}) with the FD distribution for various values of the
added energy.

When the system is in contact with a heat bath at temperature
reciprocal $\beta$, $r$ fluctuates with a probability law $q^r$ where,
as before, $q \equiv \exp (-\beta)$.  Accordingly, occupancies are
obtained by multiplying the numerator and denominator of the previous
expression in Eq.\ (\ref{eq:B3}) by $q^r$ and summing over $r$ from 0
to $\infty$.  The denominator gives the sum over states
\begin{eqnarray}
     Z^{\star}(q) & = & \sum_{r}q^r
     \sum_{n}\sum_{r_{1}}p(r_{1})p(r-n^2-r_{1}) \nonumber \\
     & = & {\left( \sum_{r}q^r p(r) \right)}^2 \sum_{n}q^{n^2}
     \label{eq:B4}
\end{eqnarray}
which may be written as an infinite product \cite{Berndt}. The
average added energy reads
\begin{eqnarray}
     r_{c}(q) & = & \frac{q}{Z^{\star}}
     \frac{{\mathrm{d}}Z^{\star}}{{\mathrm{d}}q} \nonumber \\
     & = & \sum_{j=1,2\ldots}\left( \frac{2 j}{q^{-j}-1} + \frac{(-1)^j
     2 j}{q^j - q^{-j}} \right)
     \label{eq:B5}
\end{eqnarray}

The expression for the occupancy may be reduced to a double sum, which
coincides with the one given in Eq.\ (\ref{e8}), shifted by $\pm n$
and weighted by $q^{n^2}$.  This final result is intuitive since
unbalancing between $N^{+}$ and $N^{-}$ increments the energy by
$n^2$.  We have
\begin{equation}
     n_{c}(q;k) = \frac{\sum_{n}\left( n(q;k-n) + n(q;k+n) \right)
     q^{n^2}}{\sum_{n}q^{n^2}}
     \label{eq:B6}
\end{equation}
where the sums over $n$ run from $-\infty$ to $+\infty$, and $n(q;k)$
is given in Eq.\ (\ref{eq:A12}).  The two terms in the numerator give
the same contributions.  The occupancy $n_{c}(q;k)$ may be expressed
as a function of the average energy and $k$ with the help of Eq.\
(\ref{eq:B5}).  Comparison with the FD distribution is in Fig.\
\ref{fig:1}e.

Similarly the fluorescence is obtained by shifting $k$ by $n$ with a
weight factor $q^{n^2}$.

\onecolumn

\end{document}